\documentclass[reprint]{revtex4-1} 

\usepackage{graphicx}
\usepackage[english]{babel}
\usepackage{amsthm,amsfonts,amsmath,amssymb,amscd,wasysym}

\usepackage{upgreek}

\begin{document}
\title{Polychromatic, continuous-wave mirrorless lasing from monochromatic pumping of cesium vapor}



\author{D. Antypas} 
\affiliation{Helmholtz-Institut Mainz, Mainz 55128, Germany}
\author{O. Tretiak}
\affiliation{Johannes Gutenberg-Universit\"at Mainz, Mainz 55128, Germany}
\author{D. Budker}
\affiliation{Helmholtz-Institut Mainz, Mainz 55128, Germany}
\affiliation{Johannes Gutenberg-Universit\"at Mainz, Mainz 55128, Germany}
\affiliation{Department of Physics, University of California at Berkeley, California 94720-300, USA}
\author{A. Akulshin}
\affiliation{Swinburne University of Technology, Center for Quantum \& Optical Science, POB 218, Melbourne, Vic 3122, Australia}


\begin{abstract}
We report on studies of simultaneous continuous-wave mirrorless lasing on multiple optical transitions, realized by pumping hot cesium vapor with laser light resonant with the 6S$_{1/2}\rightarrow$ 8P$_{3/2}$ transition. The multiplicity of decay paths for the excited atoms to their ground state is responsible for the emergence of lasing in a number of transitions, observed here in at least seven wavelengths in the  infrared (IR), and at two wavelengths in the blue. We study the properties of the fields generated in the cesium vapor, such as  optical power, directionality and optical linewidth. 
\end{abstract}



\maketitle

Amplified spontaneous emission (ASE) in an optically pumped medium can give rise to directional emission of coherent light, generated without the need for feedback on the medium from optical reflectors.  The first demonstration of this mirrorless lasing (ML) \cite{Jacobs1961CoherentVapor} was done shortly after the invention of the laser, and was based on ASE in cesium (Cs) vapor which was pumped with a helium lamp. Subsequently, ML was studied in alkali vapors in pulsed \cite{Sorokin1969InfraredRb2, Sorokin1971InfraredMolecules} and continuous-wave (CW) mode \cite{Sharma1981Continuous-waveVapors}. Continuation of those early works lead to the development of optically-pumped alkali lasers \cite{Pitz2017RecentLasers}. A number of studies explored the connection and interplay between ASE and nonlinear parametric wave-mixing processes in alkali vapors, that might find applications in quantum information and communication. For instance, a scheme to implement quantum memory with interface to telecom wavelengths was developed in   \cite{Kuzmich2010AConversion}.  Photon-storage \cite{Camacho2009Four-wave-mixingVapour} and implementation of correlated photon pairs \cite{Willisetal.2011PhotonEnsemble, Gulati2014GenerationAtoms} were also reported.

Various investigations of ASE and four-wave-mixing (FWM) processes in the context of ML were reported in recent years. Generation of coherent blue light through frequency-up-conversion  was done in a series of experiments employing rubidium (Rb)  \cite{Zibrov2002EfficientMedia, Jeppesen2006BlueRubidium, Akulshin2009CoherentVapour,Vernier2010EnhancedVapor, Sulham2010BlueAbsorption,Brekke2013ParametricLaser, Sell2014CollimatedVapor} or Cs vapors \cite{Schultz2009CoherentVapor, Sulham2010BlueAbsorption}. Generation of directional ultraviolet (UV) light by  stepwise excitation to a low-lying Rydberg state in Rb was demonstrated in \cite{Lam2019CollimatedVapor}. Properties of frequency-down-converted directional IR light  were also investigated in a number of experiments \cite{Akulshin2014DirectionalVapor,Sell2014CollimatedVapor, Akulshin2018Continuous-waveVapors,Sebbag2019GenerationVapor}. The competition between ASE and FWM processes, which are responsible for the directionally emitted optical fields was studied in  \cite{Boyd1987CompetitionProcess,Akulshin2014DirectionalVapor,Sebbag2019GenerationVapor}. Transfer of orbital angular momentum from one spectral region to another using FWM was explored in \cite{Walker2012Trans-spectralVapor, Chopinaud2018HighVapor,Offer2018SpiralVapour} and a method to distinguish ASE and FWM processes using vortex light was suggested and demonstrated in  \cite{Akulshin2015DistinguishingTransfer}. Finally, investigations of backward emission linked to remote sensing applications in the atmosphere were reported in a number of works (see, for example, \cite{Dogariu2011HighAir,Kartashov2012Free-spaceFilament,Mitryukovskiy2014BackwardPulses, Akulshin2018Continuous-waveVapors}). 

The majority of experiments studying aspects of CW ML in the alkali vapors employed either step-wise or two-photon excitation of the $n$S$_{1/2}\rightarrow n$D$_{5/2}$ transition, where \textit{n}  is the principal quantum number. A noticeable exception was the work \cite{Sharma1981Continuous-waveVapors}, involving single-photon excitation of the  $n$S$_{1/2}\rightarrow (n+1)$P$_{J}$ transition, performed in Rb (\textit{n}=5) and Cs (\textit{n}=6) for values for the total electronic angular momentum \textit{J}=1/2 or 3/2. In addition, ML from pulsed, single-photon pumping of the Cs 6S$_{1/2}\rightarrow$8P$_{1/2}$ transition was demonstrated in the pioneering work  \cite{Jacobs1961CoherentVapor}, with observation of lasing at a single wavelength in the IR. However, to our knowledge, an investigation of the ML processes arising from  CW excitation to a high-lying
P-state, such as the $(n+2)$P$_J$,  has not been performed. Possible ASE processes on the number of decay channels from this high-lying state to the ground state has the potential to result in multi-color ML. The present work is motivated in part by ongoing work in our laboratory related to remote sensing with the use of backward-directed ASE \cite{Akulshin2018Continuous-waveVapors} and degenerate mirrorless lasing \cite{Papoyan2018EvidenceExperiment}.

\begin{figure}
\includegraphics[width=8.7cm]{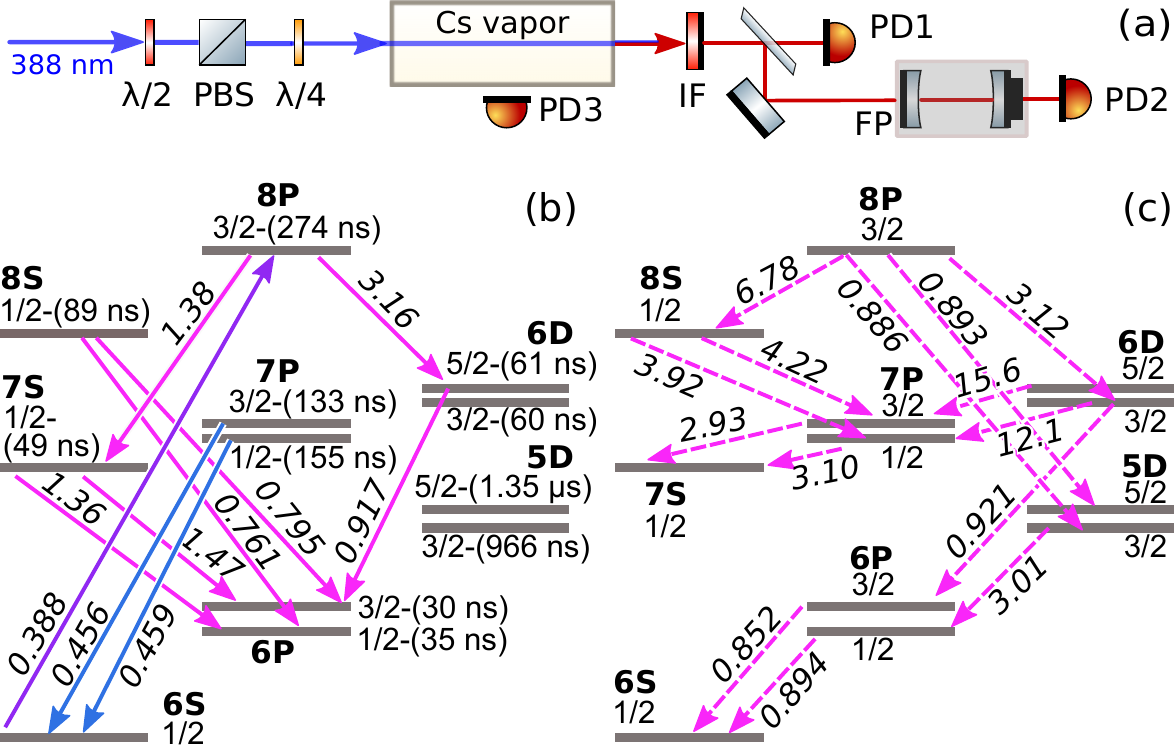}
\caption{\small{a) Simplified apparatus schematic. $\lambda/2$: half-wave plate; $\lambda/4$: quarter-wave plate; PBS: polarizing-beamsplitter; IF: interference (band-pass) filter; PD: photodetector; FP: Fabry-Perot spectrum analyzer. PD3 is used to collect fluorescence from the side of the vapor cell. b) Partial Cs energy level diagram with relevant levels. The electronic angular momentum and lifetime of the respective levels are also shown. Decays indicated by solid lines correspond to detected ML. The respective transition wavelengths are given in $\upmu$m. c) Energy level diagram that shows transitions in which ML possibly occurs, but was not detected in the present work.}  }
\label{fig:apparatus-energylevels}
\end{figure}

Creation of the required population inversion between two energy levels for directional light generation through an ASE process, is dependent upon the radiative lifetimes of the upper and lower level, as well as the probabilities for decay from the upper to the lower level ($P_1$) and from the lower level to level(s) of lower energy ($P_2$). A large ratio of upper and lower level lifetimes and a moderately small ratio $P_1/P_2$ favor this inversion.  
Here we study simultaneous multi-wavelength ML in Cs vapor, occurring when the medium is pumped with a single-wavelength CW source at 388 nm, which excites the 6S$_{1/2}\rightarrow$8P$_{3/2}$ transition. Cascade de-excitation of atoms (Fig. \ref{fig:apparatus-energylevels}b) is distributed among a number of paths in which population inversion can occur, resulting in multi-color ASE processes. 
In addition to the richness of ASE-induced lasing, this multiplicity of cascade decays, can, in principle, give rise to generation of FWM fields in a number of parametric loops. Within such loops, down-conversion in frequency of the 388 nm pump photon results in the creation of three photons, including  photons in the IR range. This might be of interest for applications in quantum information and quantum networks, in which conversion to wavelengths in the telecom IR range is beneficial.

 Observations of directional emission were made with the setup shown in Fig. \ref{fig:apparatus-energylevels}a. Cs atoms contained in a 7 cm long borosilicate-glass cell which was maintained at temperature in the range 105-140 $^{\circ}$C, were excited with up to 80 mW of 388 nm light from a frequency-doubled Ti:Sapphire laser (M$^2$ SolsTiS \& ECD-X), tuned to excite the 6S$_{1/2}$, $F=4\rightarrow$ 8P$_{3/2}$, $F'$ transition. The total angular momenta $F$ and $F'$ refer to the ground and excited state hyperfine level, respectively. The UV beam was slightly converging inside the cell, with a mean radius of \mbox{$\approx200$ $\upmu$m} ($1/e^2$ intensity) at the cell center. The cell windows allow for transmission in the 0.2-3.6 $\upmu$m range, with a maximum of $\approx$90\% between 0.2 and 2 $\upmu$m and a minimum  of 10\% at \mbox{2.5 $\upmu$m}. Directional light was detected in the forward direction with amplified Si, PbS or PbSe detectors, which operate in the 0.3-1.1, 1-2.5, or 1.5-4.8 $\upmu$m range, respectively. Long-pass as well as interference filters were used to block the residual pump light and allow for wavelength selectivity. To measure the small optical powers of some of the generated fields, lock-in detection was used in some cases. In those cases, the pump beam was chopped at a frequency of a few hundred Hz.

\begin{figure}
\includegraphics[width=8.7cm]{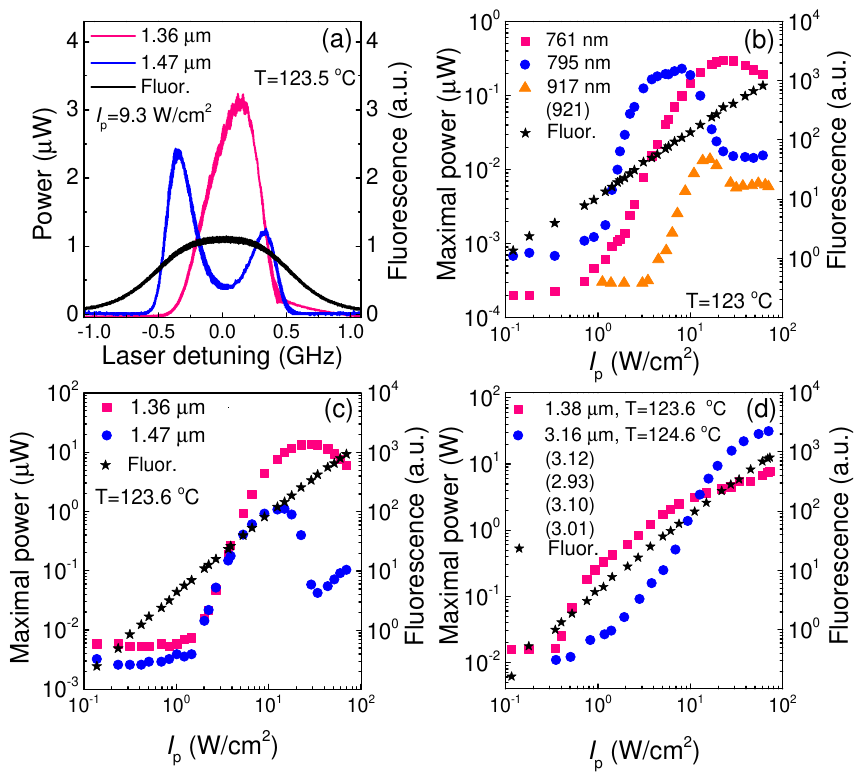}
\caption{\small{a) Power of directional light at 1.36 and 1.47 $\upmu$m and combined fluorescence signal (UV, visible and IR) collected from the side of the cell as a function of the 388 nm laser frequency detuning from the peak of the Doppler-broadened UV transition. The maximal power in such spectra is measured for different pump intensities $I_p$, and plotted in b), c), and d) to illustrate threshold behavior at the respective wavelengths. The wavelengths in parentheses indicate potential contributions to the measured signals from spectral component(s) adjacent to the primary one, due to filter imperfections.}}
\label{fig:spectrum-power}
\end{figure}

Emergence of ML between particular Cs levels occurs when population inversion between these levels is achieved, and is dependent upon parameters such as the atomic density, optical pump intensity $I_p$, polarization and frequency. In the present system, the onset of directional emission occurs at a vapor temperature greater than 110 $^{\circ}$C, or at a Cs density greater than 2.7$\times10^{13}$ cm$^{-3}$. Circular polarization for the pump beam was found to provide lower lasing threshold for all the observed IR emission lines. When ML occurred, it could be observed for a pump frequency range which is a fraction of the $\approx$1 GHz-wide Doppler-broadened  6S$_{1/2}$, $F=4\rightarrow$ 8P$_{3/2}$, $F'$ transition, whose spectrum was observed via fluorescence collected with a Si photodiode from the side of the cell (Fig. \ref{fig:apparatus-energylevels}a). Figure \ref{fig:spectrum-power}a shows directional emission at 1.36 and 1.47 $\upmu$m and fluorescence (UV, visible and IR), recorded by sweeping the pump laser frequency. Unlike the fluorescence level, which grows linearly with the intensity $I_p$, directed emission exhibits threshold behavior. Such a behavior is shown in Fig. \ref{fig:spectrum-power}b, Fig. \ref{fig:spectrum-power}c and Fig. \ref{fig:spectrum-power}d, in which the maximal power observed in spectra like that of Fig. \ref{fig:spectrum-power}a is plotted against $I_p$ for six wavelengths in the IR, for which directional light was detected. As seen in  Fig. \ref{fig:spectrum-power}b, Fig. \ref{fig:spectrum-power}c and Fig. \ref{fig:spectrum-power}d, the greatest emitted power observed was on the order of tens of $\upmu$W for 1.36  $\upmu$m and the mid-IR light around 3 $\upmu$m, while the lowest was $\approx$ 10 nW for the 917 nm emission.

The finite band-pass width of the interference filters used may, in particular cases, have resulted in parasitic contributions to measurements from spectral components adjacent to the desired wavelengths. One such case involves detection of 3.16 $\upmu$m light from the $8P_{3/2}\rightarrow 6D_{5/2}$ transition, which is likely to contain contributions from at least some of the 3.12, 2.93,  3.10 or \mbox{3.01 $\upmu$m} components (see Fig. \ref{fig:apparatus-energylevels}c), as the filter used for this measurement has a 500 nm FWHM. A similar measurement imperfection may have occurred in the detection of 917 nm light, which is too close spectrally to the 921 nm component for our 12 nm-wide filter to isolate. However, inspection of the 917 nm emission with a Fabry-Perot (FP) spectrum analyzer did not reveal traces of directional emission at 921 nm, which points to 921 nm being either weak or absent.

A noticeable aspect to some of the observed ML processes, is the apparent competition within certain pairs of transitions. This competition occurs because those transitions share an excited state, and  it was seen between lasing at 795 and 761 nm, both related to decay from the 8S$_{1/2}$ state, and between the 1.47 and 1.36 $\upmu$m fields, both generated through decay from the 7S$_{1/2}$ state. At low pump intensity, emission at 795 nm is dominant compared to that at 761 nm (Fig. \ref{fig:spectrum-power}b), while lasing at both 1.36 and 1.47 $\upmu$m is of approximately equal power (Fig. \ref{fig:spectrum-power}c). Lasing in the low- pump-intensity regime is observed to occur for all wavelengths within a pump frequency region  which is centered around the peak of the Doppler-broadened 388 nm transition. As pump intensity increases, the 761 nm and 1.36 $\upmu$m components initiate a competition with the other wavelength of the pair. As a result, with further increase of pump intensity, the 795 nm and 1.47 $\upmu$m components are gradually limited to lasing away from the center of the velocity distribution of the pump transition. The onset of this process is shown in  Fig. \ref{fig:spectrum-power}a for the 1.47 $\upmu$m field, and as a result of the competition, the power of the 795 nm and 1.47 $\upmu$m fields exhibits a dramatic decrease, as seen in Fig. \ref{fig:spectrum-power}b and Fig. \ref{fig:spectrum-power}c, respectively. 

\begin{figure}
\includegraphics[width=8.7cm]{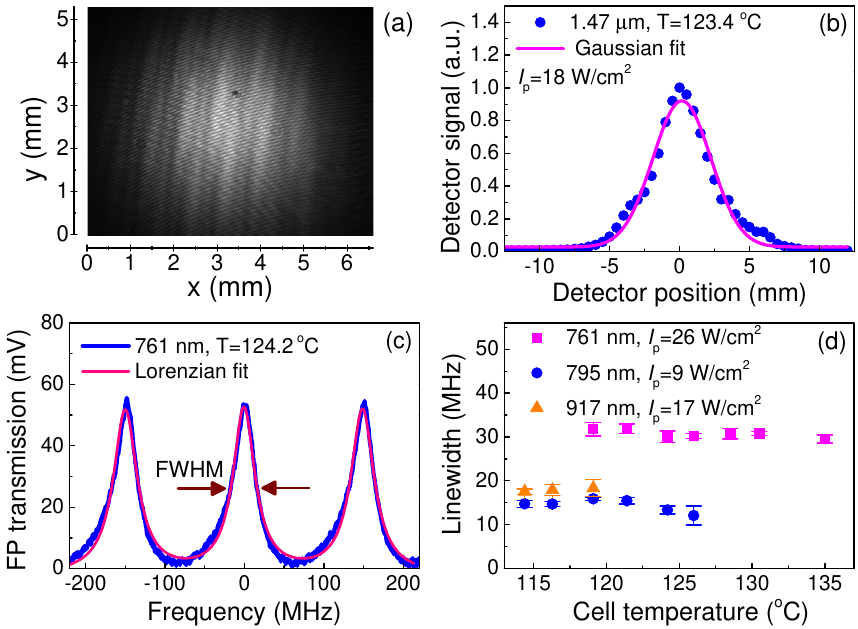}
\caption{\small{Directionality and linewidth of selected ML components. a) Image of the 795 nm beam, taken with a CMOS camera of active area 6.6$\times$5.3 mm$^2$, located at a distance of 44 cm from the cell center. The observed interference fringes are due to imperfections in the filter used to select the particular wavelength \cite{Sushkov2005OnFilters}. b) Profile of the 1.47 $\upmu$m beam, recorded via transverse translation of a photodetector fitted with a 1.2 mm-wide aperture, and positioned 80 cm away from the cell center. c) Spectrum of the 761 nm light, measured with a FP. Analysis of such spectra provides estimates for the linewidths of the 761, 795 and 921 nm beams, shown  in d). These linewidths were determined within temperature ranges for which the emitted power in the respective wavelengths was large enough to obtain a FP spectrum.}}
\label{fig:profile-linewidth}
\end{figure}

Directionality of the emitted IR light was checked with the use of different methods, depending on the particular wavelength. For the 761, 795 and 917 nm components, a CMOS camera was used to record beam profiles at several positions \textit{x} in the path of the beams (Fig. \ref{fig:profile-linewidth}a), allowing to determine their divergence. A half-angle divergence $\Delta\theta_{1/2}\equiv \Delta r/\Delta x$, was determined for each beam, where \textit{r} is the $1/e^2$ intensity radius of the gaussian fit to the respective beam profiles.  The results for the 761, 795 and 917 nm beams are $\Delta\theta_{1/2}=$5.2(6), 5.7(7) and 5.4(10) mrad, respectively.  The profiles of the 1.36, 1.38, 1.47 and 3.16 $\upmu$m beams were checked by measuring light power with a detector whose position was stepped transversely to the beam direction (Fig. \ref{fig:profile-linewidth}b). The respective results are $\Delta\theta_{1/2} \approx$ 
6, 8, 5 and 12 mrad, with an estimated error of 1 mrad.  The divergence and diffraction of all the observed IR components was found to be consistent to within a factor of two with the expectation of $\Delta\theta_{1/2}\approx 2w/L=5.7$  mrad, for light generated through an ASE process within a pencil-shaped gain region, of approximate diameter $2w=400$ $\upmu$m and length $L$=7 cm \cite{Peters1972AmplifiedCoherence}.

The linewidth of the 761, 795 and 917 nm fields was evaluated with use of a FP spectrum analyzer, which has a finesse of $\approx$30 and an intrinsic FWHM linewidth of $\approx$5 MHz within the spectral range of interest. Recorded spectra such as that shown in Fig. \ref{fig:profile-linewidth}c, were deconvolved to obtain the contribution of the respective linewidths to the FP fringes. These results are shown in Fig.  \ref{fig:profile-linewidth}d  and represent data taken at several vapor temperatures. The linewidth dependence on temperature for the near-IR lasing components is not pronounced.

In addition to the multitude of the observed IR components, blue light related to the 7P$_{J} \rightarrow$ 6S$_{1/2}$ transitions (with wavelengths 459 nm for J=1/2 and 456 nm for J= 3/2) was also detected. Directional CW blue light generation at 455 nm was reported in \cite{Schultz2009CoherentVapor} using a step-wise  excitation scheme. Simultaneous dual-wavelength blue light generation at 455 and 459 nm was realized in \cite{Sulham2010BlueAbsorption} using a pulsed-laser pumping scheme. To our knowledge, the latter observations have not been previously reported in the CW regime. This example of frequency down-conversion can, in principle, arise, for each of the two blue colors, within two different parametric processes: the 6S$_{1/2}\rightarrow 8$P$_{3/2}\rightarrow $ 8S$_{1/2}\rightarrow$ 7P$_{J}\rightarrow $6S$_{1/2}$ and the 6S$_{1/2}\rightarrow 8$P$_{3/2}\rightarrow $ 6D$_{J'}\rightarrow$ 7P$_{J}\rightarrow $6S$_{1/2}$ loop (see Fig. \ref{fig:apparatus-energylevels}b), with $J'=3/2$ or $5/2$. Assuming blue-light generation occurs through FWM, each process involves mixing of the applied 388 nm field with three internally generated fields: two IR fields and the blue field. To satisfy phase-matching conditions in the case of FWM, the fields to be generated ought to propagate approximately collinearly  with the pump light.

\begin{figure}
\includegraphics[width=8.7cm]{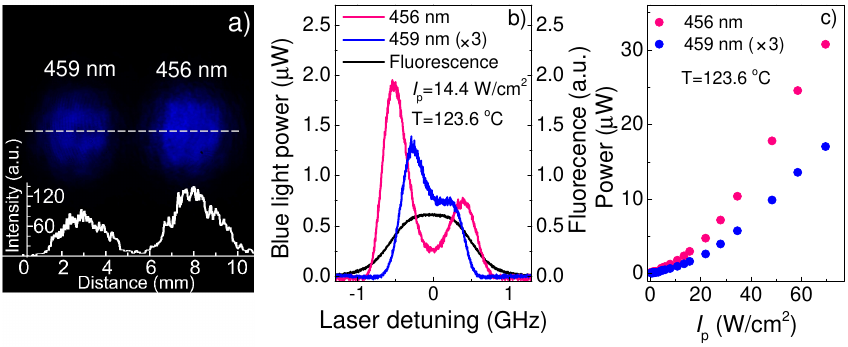}
\caption{\small{ a) Image of the 456 and 459 nm beams, recorded with a CMOS sensor. The two beams were weakly focused onto the sensor. Also shown is an intensity profile along the white dashed line. b) Spectral profile of the 456 and 459 nm directional light and fluorescence from the Cs cell. c) Power of the two blue beams, recorded for varying intensity of the pump, measured at a cell temperature of 123.6 $^\circ$C. } }
\label{fig:BL}
\end{figure}

Separating a directional blue beam using a 1200 lines/mm diffraction grating from IR lasing and residual pump light,  and observing its cross section, we found that the blue light consists of two spectral components, as shown in Fig. \ref{fig:BL}a.
The wavelengths of the two beams were measured with a HighFinesse WSU-2 wavemeter and were found to be consistent to within $\approx10^{-3}$ nm with the literature values \cite{Kramida2018NISTDatabase}. Beam profiling at several positions was done to measure the respective divergences, with the result of  \mbox{$\Delta\theta_{1/2}$=1.6(2) mrad} for the 456 and  $\Delta\theta_{1/2}$=1.5(2) mrad for the \mbox{459 nm} light. These values are roughly consistent with the diffraction limit of $\lambda/\pi w\approx0.7$ mrad, for a gaussian beam whose waist at the focus is equal to the $\approx$ 200 $\upmu$m waist of the gain region in our setup. 

As the generated blue light is related to the relatively strong 6S$\rightarrow$7P$_{J}$ transitions, it is partially absorbed by the Cs vapor. For moderate pump intensity, strongest emission with respect to pump frequency therefore, tends to appear away from the center of of the velocity distribution of Cs atoms, as shown in Fig. \ref{fig:BL}b. The \mbox{120-125 $^{\circ}$C} temperature range for the Cs vapor was found to result in the strongest emission. Below this range, decreased Cs density yields weaker emission, while above it, absorption of blue light becomes strong. Figure \ref{fig:BL}c shows blue light power generated as a function of  pump intensity $I_p$, at a temperature close to the optimal. The highest power generated is $\approx$30 $\upmu$W for 456 nm and  $\approx$6 $\upmu$W for 459 nm. The pump laser polarization that maximized conversion efficiency to the blue was found to be approximately linear. Unlike ASE IR generation, in the blue light generation, no saturation is observed, even at the highest pump power of $\approx$ 80 mW. 

Further experiments are required to characterize the nature of the blue light generation (i.e. ASE- or FWM-induced) as well as the respective contributions of the two parametric processes which are involved in the generation of each blue wavelength. Study of orbital angular momentum transfer, which is expected to occur in a parametric FWM process  \cite{Akulshin2015DistinguishingTransfer}, or a check for correlations between spectra of the blue and IR fields that are involved in the same loop (as seen in Refs. \cite{Akulshin2014DirectionalVapor, Sebbag2019GenerationVapor}), would be  possibilities for further explorations. 

Despite strong fluorescence observed at 852 nm, directional emission at this wavelength as a result of a FWM process was not observed. Such a field is at the frequency of the strong D2 line, and if generated, is absorbed by the dense Cs vapor.

In conclusion, we have implemented a scheme to induce simultaneous CW mirrorless lasing in a number of colors, by exciting a high-lying P-state in Cs vapor with laser light at \mbox{388 nm}. The generated power and directionality of all the observed fields was measured, as well as the spectral linewidth of some of those fields. Directional emission was detected in a total of nine wavelengths, including two fields at the blue, created within distinct parametric processes. The mirrorless lasing studied here was co-propagating with the pump laser light, however, lasing due to ASE is not necessarily equally intense in the co- and counter-propagating directions, as was shown in \cite{Akulshin2014DirectionalVapor}. Further studies of backward-directed ASE that are potentially important for remote sensing  will be pursued.

The simplicity of the experimental setup, which requires a single pump laser of relatively low power (tens of mW) at a wavelength covered by commercially available near-UV laser diodes, might make the system under investigation attractive for applications that make use of correlated photon pairs in the IR range. Another possible application could be in the generation of mid-IR light for spectroscopy, as mirrorless lasing was observed in seven IR transitions, with more, currently undetected components likely present in the emission spectrum.  One such example is the 6.78 $\upmu$m field, which in the present setup is absorbed by the vapor cell windows. Fields at 4.2 and \mbox{3.9 $\upmu$m} from the 8S$\rightarrow$7P decays, as well as 15.6 and \mbox{12.1 $\upmu$m} from the 6D$\rightarrow$7P decays are additional examples. CW population inversion on the 6D$\rightarrow$7P transitions is still possible despite unfavorable lifetimes of the upper and lower levels, as the probability of all transitions from 6D levels to 7P levels is smaller than the probability of decays from the 7P levels \cite{Heavens1961RadiativeMetals}. \\

\noindent \textbf{Funding}
This work was supported in part by ONRG, grant number N62909-16-1-2113. 

\noindent \textbf{Acknowledgments}  
We acknowledge fruitful discussions with A. Wickenbrock,  F. Pedreros Bustos and N. Rahaman.\\

\bibliography{references}



\end{document}